\documentclass[11pt,tightenlines]{revtex4}
\usepackage{amsfonts}
\usepackage{amsmath}
\usepackage{txfonts}
\usepackage{graphicx}
\usepackage{amssymb}
\usepackage{color}
\usepackage{wrapfig,lipsum,booktabs}
\usepackage{floatrow}
\usepackage{epstopdf}
\usepackage{hyperref}
\usepackage[normalem]{ulem}

\begin{document}
	\title{Through synapses to spatial memory maps: a topological model}
	\author{Yuri Dabaghian}
	\affiliation{Department of Neurology, The University of Texas McGovern Medical School, 6431 Fannin St, Houston, TX 77030\\
		e-mail: dabaghian@gmail.com}
	\date{\today}

\begin{abstract}
\textbf{Abstract}. Various neurophysiological and cognitive functions are based on transferring information between 
spiking neurons via a complex system of synaptic connections. In particular, the capacity of presynaptic inputs to 
influence the postsynaptic outputs---the efficacy of the synapses---plays a principal role in all aspects of hippocampal 
neurophysiology. However, a direct link between the information processed at the level of individual synapses 
and the animal's ability to form memories at the organismal level has not yet been fully understood. Here, we 
investigate the effect of synaptic transmission probabilities on the ability of the hippocampal place cell 
ensembles to produce a cognitive map of the environment. 
Using methods from algebraic topology, we find that weakening synaptic connections increase spatial learning 
times, produce topological defects in the large-scale representation of the ambient space and restrict the range 
of parameters for which place cell ensembles are capable of producing a map with correct topological structure.
On the other hand, the results indicate a possibility of compensatory phenomena, namely that spatial learning 
deficiencies may be mitigated through enhancement of neuronal activity.
\end{abstract}

\maketitle
\newpage

\section{Introduction}
\label{section:intro}

The location-specific spiking activity of the hippocampal neurons, known as place cells \cite{OKeefe}, gives rise 
to an internalized representation of space---a cognitive map. Each place cell fires a series of action potentials in 
specific spatial region---its place field, so that the ensemble of such cells produces a ``map" of the environment 
in which they are active (Fig.~\ref{Fig1}A). By construction, such a map defines the temporal order in which place 
cells fire as the animal explores the environment, and therefore it can be viewed as a geometric representation of 
the spatial memory framework encoded by the hippocampus \cite{Moser,Schmidt}. 

The exact nature of this framework is currently actively studied both computationally and experimentally. 
For example, it was demonstrated that if the shape of the environment gradually changes, then the place field map 
deforms in a way that preserves mutual overlaps, adjacencies, containments, etc., between the place fields 
\cite{Gothard,Leutgeb,Wills,Touretzky,eLife}. This observation implies that the sequence in which the place 
cells fire during animal's navigation remains invariant throughout the reshaping of the arena and suggests that the 
place cells do not represent precise geometric information, but a set of qualitative connections between portions of 
the environment---a topological map \cite{Poucet,Alvernhe1,eLife}.

From the computational perspective, the topological nature of the cognitive map suggests that the information 
transmitted via place cell spiking should be amenable to topological analyses. In our previous work 
\cite{PLoS,Arai,Basso,Hoffman,CAs}, we developed a topological model that allows tracing how the information 
provided by the individual place cells may combine into a large-scale topological map of the navigated space and 
quantifying the contributions of different neurophysiological parameters. However, previous studies did not include a 
key physiological aspect---the contribution of the synaptic connections into the processes of assembling the map. 
Below we will use the topological approach to model how synaptic imperfections can affect the topological structure 
of the cognitive map, its dynamics and its stability.

The paper is organized as follows. First, we outline the basic ideas and the key concepts used in the topological 
model---simplexes, simplicial complexes, topological loops, Betti numbers, etc., and explain how these concepts 
can be applied for describing hippocampal physiology. Second, we outline the parameters of synaptic connectivity 
and the constructions used to incorporate these parameters into the model. The analyses of the outcomes is given 
in the Results section and their implications are outlined in the Discussion.

\section{The Model}
\label{section:model}

\begin{figure} 
	\includegraphics[scale=0.82]{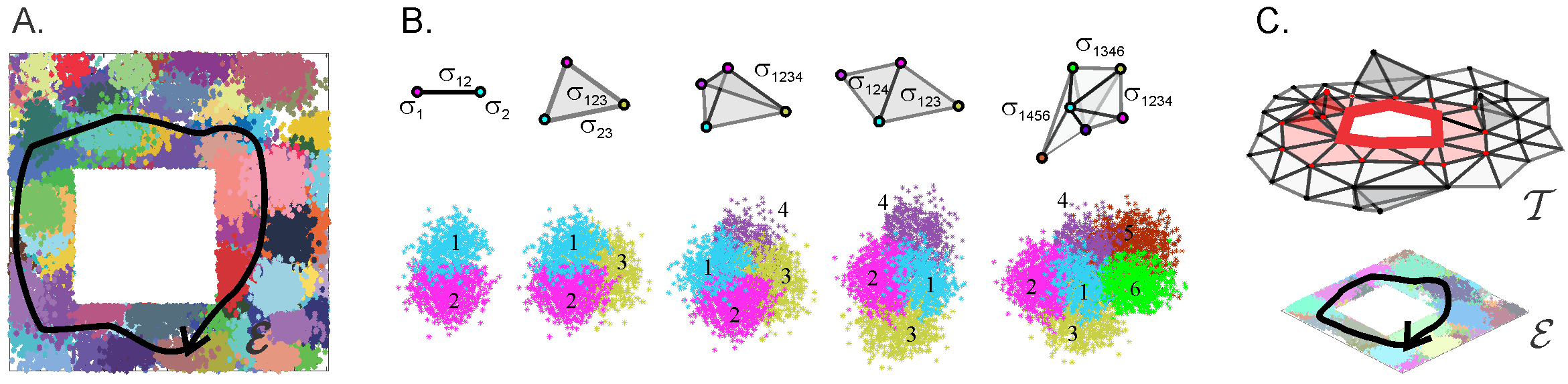}
	\caption{\label{Fig1} 
{\footnotesize
	\textbf{Place field map and nerve complex}. ({\bf A}). A place field map in a small $1m \times 1m$ environment 
		with one hole: spikes produced by different place cells are marked by dots of different colors. ({\bf B}). 
		In a schematic description of the place field map, each place field center gives rise to be a zero-dimensional 
		vertex ($0D$ simplex $\sigma_{i}$); each pair of the overlapping place fields is represented by a link between 
		corresponding vertices ($1D$ simplex $\sigma_{ij}$); a triple of overlapping place fields by a triangle ($2D$ 
		simplex $\sigma_{ijk}$), four simultaneously overlapping place fields are represented by a solid tetrahedron, 
		($3D$ simplex $\sigma_{ijkl}$) etc. A less dense place field map is represented by two adjacent triangles---a 
		simple example of a nerve complex. A place field map that consists of six place fields is represented by a nerve 
		complex that consists of three tetrahedrons, $\sigma_{1234}$, $\sigma_{1456}$ and $\sigma_{1346}$. ({\bf C}). 
		According to the Alexandrov-\v{C}ech's theorem, the nerve complex construction for a place field map has the 
		same topological shape as the underlying environment---in case of the map shown on panel A, the nerve complex 
		$\mathcal{N}$ has one connected piece and contains a hole in the middle.}
}
\end{figure}

\textbf{Topological description of the place cell spiking patterns}. It is generally believed that the information 
encoded by the place cell network is represented by the connectivity between the place fields. A specific link is
suggested by the classical Alexandrov-\v{C}ech's theorem of Algebraic Topology asserts that the pattern of overlaps 
between regions that cover a space $X$ does, in fact, capture its topological structure \cite{Alexandroff,Cech}. 
The implementation of this theorem is based on constructing the so-called ``nerve simplicial complex" $\mathcal{N}$, 
whose vertexes correspond to the individual domains of the cover: one-dimensional ($1D$) links---to their pairwise 
overlaps, two-dimensional ($2D$) facets---to the triple overlaps and so forth  (Fig.~\ref{Fig1}B). In other words, 
each $n^{th}$ order overlap between the place fields is schematically represented by an $n$-dimensional simplex 
$\sigma$, so that the full set $\mathcal{N}$ of such simplexes incorporates the connectivity structure of the entire 
place field map \cite{Ghrist,Curto,PLoS}. According to the Alexandrov-\v{C}ech's theorem, this complex has the same 
``topological shape" as $X$, i.e., the same number of pieces, gaps and holes \cite{Hatcher,Alexandrov}, which provides 
a link between the place cells' spiking pattern and the topology of ambient space \cite{PLoS,Arai,Basso,Hoffman,CAs}, 
exploited below.

%%%%%%%%%%%%%%%%%%%%%%%%%%%%%%%%%%%%%%%
%%%%%%%%%%%%%%%%%%%%%%%%%%%%%%%%%%%%%%%

In general, simplicial complexes provide a convenient framework for describing a wide scope of physiological phenomena. 
For example, the combinations of the place fields traversed during the rat's moves correspond to a chain of simplexes 
$\Gamma =\{\sigma_{1},\sigma_{2},\ldots,\sigma_{k}\}$ that qualitatively represents the shape of the physical trajectory: 
a closed chain represents a closed physical route, a pair of topologically equivalent chains represent two similar physical 
paths and so forth \cite{Guger,Brown1}. The pool of such chains can be used to describe the topological shape of the entire 
complex---and hence of the corresponding environment. For example, the number of chains that can be deformed into the 
same vertex defines how many disconnected pieces $\mathcal{N}$ has. The number of topologically inequivalent chains that 
contract to a closed sequence of links defines the number of distinct holes that prevent these chains from contracting to 
vertexes and so forth \cite{Hatcher,Alexandrov}. In the following, we will refer to these two types of chains, counted up 
to topological equivalence, as to zero-dimensional ($0D$) and one-dimensional ($1D$) ``topological loops'' (a standard 
mathematical terminology), evaluate their numbers---in mathematical terms, zeroth and first Betti numbers, $b_0(\mathcal{N})$ 
and $b_1(\mathcal{N})$, and use them to describe shapes of the simplicial complexes.

\begin{figure}[!ht]
\includegraphics[scale=0.84]{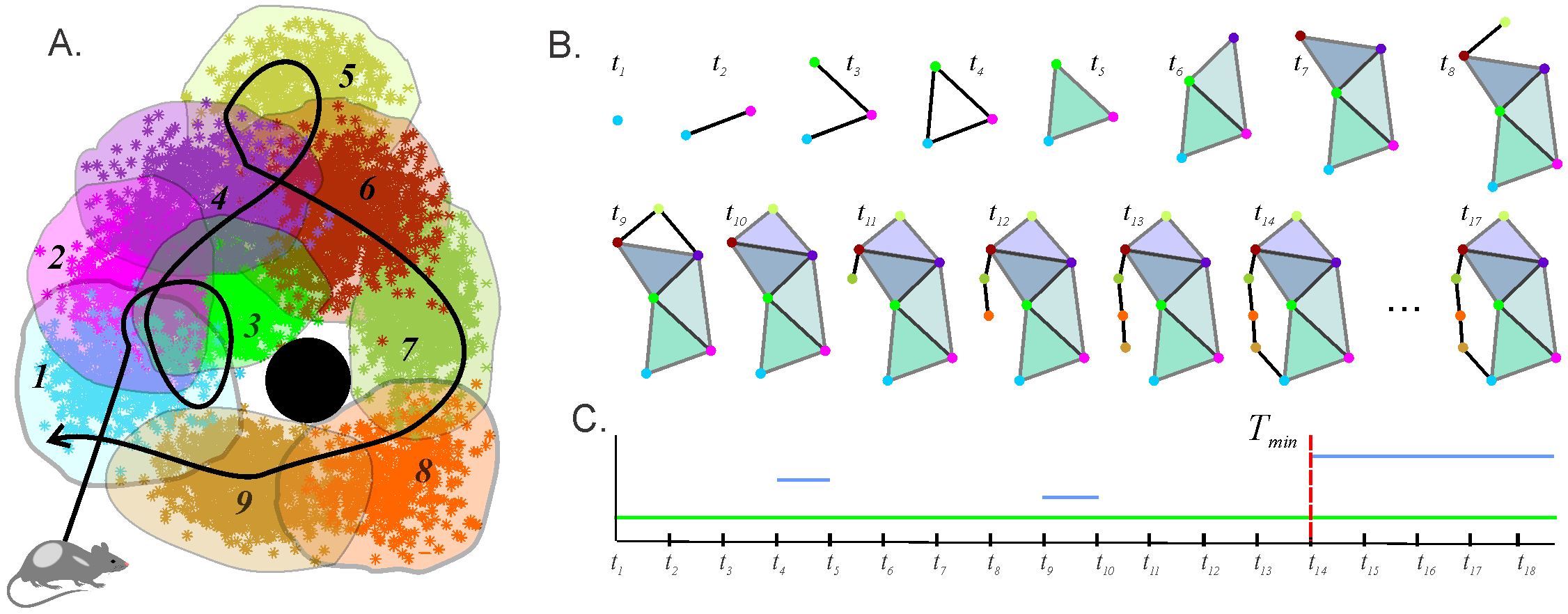}
\caption{\label{Fig2}
{\footnotesize
	\textbf{Figure2. The dynamics of the topological information.}
		({\bf A}). A mini place field map: nine place fields, enumerated in sequence they are traversed by the 
		animal's trajectory (black curve). The black circle in the middle represents an obstacle. First the animal 
		enters the place field 1 at a moment, $t_{1}$, the nerve complex $\mathcal{N}(t_{1})$ shown on the 
		panel ({\bf B}) acquires a vertex $\sigma_{1}$ (blue dot). At the time $t_{2}$, the animal crosses the 
		domain where the blue and the magenta place fields overlap, and the nerve complex $\mathcal{N}(t_2)$ 
		acquires the vertex $\sigma_{2}$ and the edge $\sigma_{12}$ between these two vertices. Then the animal 
		enters the place field 3, which contributes a vertex $\sigma_3$ and a link $\sigma_{23}$ to $\mathcal{N}(t_3)$.
		As the trajectory goes back to the first place field, the complex $\mathcal{N}(t_{4})$ acquires a loop. 
		At the moment $t_{5}$ the animal gets into the region where three place fields (1, 2 and 3) overlap; as a
		result, a two-dimensional simplex $\sigma_{123}$ appears in $\mathcal{N}(t_{5})$ and closes the loop. 
		At time $t_{6}$ the animal gets into the intersection of place fields 4, 5 and 6, which contributes the second 
		filled triangle to $\mathcal{N}(t_{6})$, and so on. At the moment $t_{11}$ the animal's trajectory starts to 
		go around the obstacle, and the nerve complex begins to grow a handle which closes into a loop at $t_{14}$. 
		After the animal has probed all intersection domains, the structure of the nerve complex ceases to change. 
		({\bf C}). Each horizontal bar represents the timeline of a $0D$ or a $1D$ loop in $\mathcal{N}(t)$. Notice, 
		that there is only one persistent $0D$ loop, because, at all times, there is only one connected piece in 
		$\mathcal{N}(t)$. In addition, there are three $1D$ loops: two of them are spurious, appearing at $t_{4}$ 
		and at $t_{9}$ and disappearing in one time step. In contrast, the loop that appeared at $t_{14}$, after 
		all the place fields and their intersections are visited, persists forever and thus represents stable 
		topological information. The time $T_{\min} =  t_{14}$ thus provides an estimate for the time required to 
		``learn" this particular map.}
}
\end{figure}

\textbf{Learning dynamics}. To describe how the animal ``learns" the environment, one can follow how the nerve complex and 
its Betti numbers develop in time. In the beginning of exploration, the nerve complex represents connections between the place 
fields that the animal had time to visit. Such a complex is small and may contain gaps that do not necessarily correspond to 
physical holes or inaccessible spatial domains of the environment. As the animal continues to navigate, the nerve complex grows 
and acquires more details; as a result, its the spurious gaps and holes (topological noise) disappear, leaving behind a few 
\emph{persistent} ones that represent stable topological information (Fig.~\ref{Fig2}). The minimal time, $T_{\min}(\mathcal{N})$, 
required to recover the correct number of topological loops,
\begin{equation}
T_{\min}(\mathcal{N}): \,\,\,b_k(\mathcal{N},t)=b_k(\mathcal{E})\,\,\,\textrm{for}\,\,\, t>T_{\min}(\mathcal{N}) \,\,\,\textrm{and}\,\,\,k\geq 0,
\label{tmin}
\end{equation}
can be used as a theoretical estimate of the time needed to learn path connectivity \cite{PLoS}. In the case of the environment 
illustrated on Fig.~\ref{Fig1}A, with the Betti numbers $b_0(\mathcal{E})=b_1(\mathcal{E})=1$, $b_{k>1}(\mathcal{E})=0$, the nerve
complex is expected to have the same  ``topological barcode": $b_0(\mathcal{N},t>T_{\min})=b_1(\mathcal{N},t>T_{\min})=1$, $b_{k>1}
(\mathcal{N},t>T_{\min})=0$.

%%%%%%%%%%%%%%%%%%%%%%%%%%%%%%%%%%%%%%%
%%%%%%%%%%%%%%%%%%%%%%%%%%%%%%%%%%%%%%%

\textbf{Temporal coactivity complex}. From the physiological perspective, the arguments based on the analyses 
of place fields and trajectories provide only an indirect description of information processing in the brain. 
In reality, the hippocampus and the downstream brain regions do not have access to the shapes and the locations 
of the place fields or to other artificial geometric constructs used by experimentalists to visualize their data. 
Physiologically, the information is represented via neuronal spiking activity: if the animal enters a location 
where several place fields overlap, then there is a \emph{probability}, modulated by the rat's location, that the 
corresponding place cells will produce spike trains that overlap \emph{temporally}. This pattern of coactivity 
signals to the downstream brain areas that the regions encoded by these place cells overlap. Thus, in order to 
describe the learning process in proper terms, one needs to construct a \emph{temporal} analogue of the nerve 
complex based only on the spiking signals, which is, in fact, straightforward. Indeed, one can represent an active 
place cell, $c_{i}$, by a vertex $v_{i}$; a pair of coactive place cells, $c_{i}$ and $c_{j}$---by a bond $\sigma_{ij}$ 
between the vertices $v_{i}$ and $v_{j}$; a coactive triple of place cells, $c_{i}$, $c_{j}$ and $c_{k}$---by a 
three vertex simplex $\sigma_{ijk}$ and so on \cite{Ghrist,Curto,PLoS}. This construction produces a time-dependent 
``coactivity complex" $\mathcal{T}(t)$---a temporal analogue of the nerve complex $\mathcal{N}(t)$ constructed above, 
whose dynamics can also be used to model topological learning, e.g., to compute the learning time from the spiking 
data, $T_{\min}(\mathcal{T})$, and so forth \cite{PLoS}. 

\textbf{Cell assembly complex}.  The construction of a temporal complex can be refined to reflect more subtle 
physiological details, e.g., the functional organization of the hippocampal network. Studies of place cells' 
spiking times point out that these neurons tend to fire in ``assemblies"---functionally interconnected groups 
that are believed to synaptically drive a population of ``readout" neurons in the downstream networks 
\cite{Harris1,Harris2,Jackson,ONeill,Syntax}. The latter are wired to integrate spiking inputs from their 
respective cell assemblies and actualize the connectivity relationships between the regions encoded by the 
corresponding place cells \cite{Syntax,SchemaS}.

\begin{wrapfigure}{c}{0.5\textwidth}
	\includegraphics[scale=0.7]{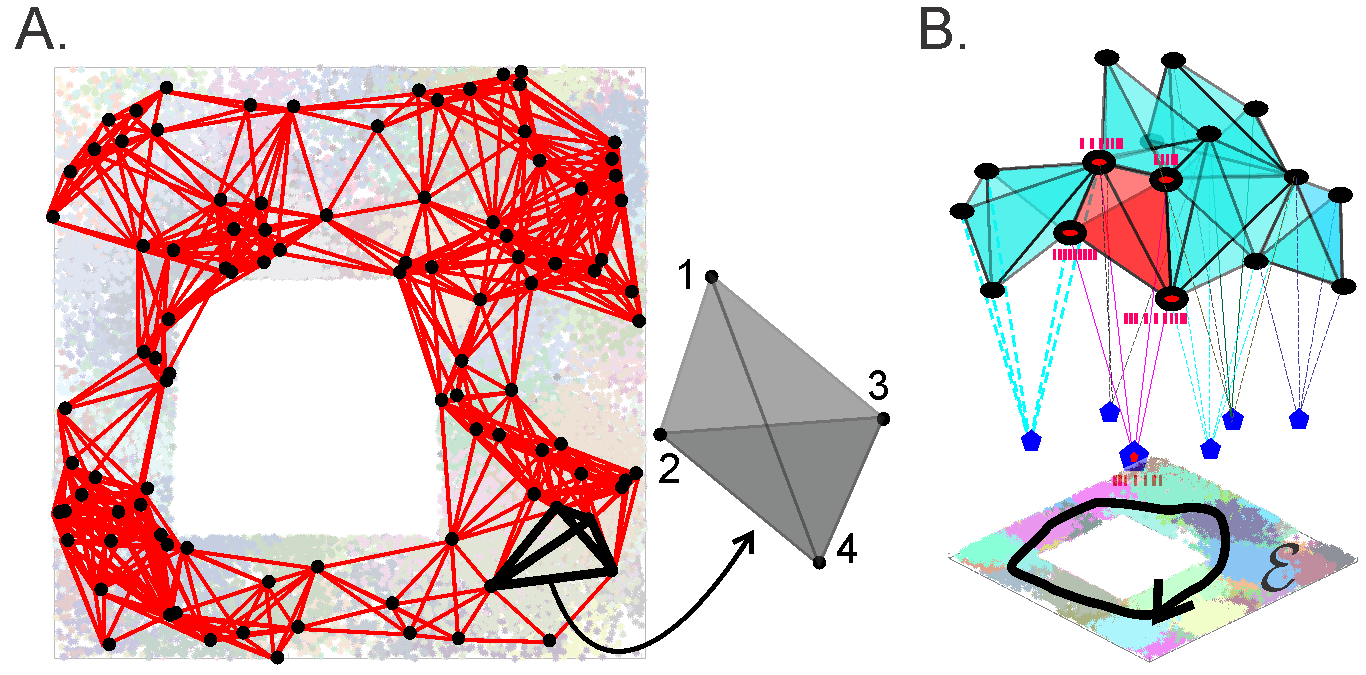}
	\caption{\label{Fig3} 
		{\footnotesize
		\textbf{Coactivity graphs and cell assembly complexes}. ({\bf A}). Active place cells are represented 
		by the vertexes of the coactivity graph $\mathcal{G}$ (black dots placed at the centers of the 
		corresponding place fields). Two vertexes are connected by an edge if the corresponding place cells
		exhibit coactivity. The fully connected subgraphs of $\mathcal{G}$---its cliques, e.g., the four 
		interconnected black links on the right---correspond to the cell assemblies. 
		({\bf B}). The collection of cliques viewed as simplexes of the cell assembly complex $\mathcal{T}_{CA}$, 
		represent the topology of underlying environment. In the model, every place cell in an assembly is 
		synaptically connected to a readout neuron (blue pentagons). The spikes from an active place cells $c_k$ 
		(an ignited cell assembly $\varsigma$ is shown in red) transmit to a readout neuron with probability $p_{k} 
		< 1$ and the readout neuron responds with probability $q_{\varsigma} < 1$.}
	}
\end{wrapfigure}

This structure can be represented by the cell assembly complex, $\mathcal{T}_{CA}$---a temporal coactivity 
complex whose maximal simplexes represent cell assemblies, rather than arbitrary combinations of coactive 
place cells.  A convenient implementation of this construction is based on the classical ``cognitive graph" 
model, in which place cells $c_{i}$ are represented as vertexes $v_{i}$ of a graph $\mathcal{G}$, while the 
connections (functional or physiological) between pairs of coactive cells are represented by the links, 
$\sigma_{ij} = [v_{i}, v_{j}]$ of $\mathcal{G}$ \cite{SchemaS,Burgess,Muller}. The place cell assemblies 
$\sigma = [c_{1}, c_{2}, \ldots, c_{n}]$ then correspond to fully interconnected subgraphs of $\mathcal{G}$, 
i.e., to its maximal cliques \cite{Hoffman,CAs}. Since a clique $\sigma$, as a combinatorial object, can be 
viewed as a simplex span by the same sets of vertexes, the collection of cliques of the coactivity graph 
$\mathcal{G}$ produces a so-called clique simplicial complex \cite{Jonsson}, which represents the population 
of place cell assemblies and may hence be viewed as a cell assembly complex $\mathcal{T}_{CA}$ (Fig.~\ref{Fig3}). 

\textbf{Phenomenological description of the synaptic parameters}. In the previous studies, we demonstrated 
that such complexes can acquire a correct topological shape in a biologically plausible period of time, in 
both planar and in voluminous environments, provided that the simulated spiking parameters values fall into 
the biological range \cite{PLoS,Arai,Basso,Hoffman,CAs}. However, the organization and the dynamics of these 
complexes did not reflect the parameters of synaptic connectivity, e.g., the mechanisms of transferring, 
detecting and interpreting neuronal (co)activity in the hippocampus and in the downstream networks. To 
account for these components, the topological model requires a basic modification: a particular coactivity 
pattern should be incorporated into an \emph{effective} coactivity complex $\mathcal{T}_{\textrm{eff}}$ not 
by the virtue of being merely produced, but by the virtue of being produced, transmitted and ultimately 
detected by a readout neuron. In other words, only \emph{detected} activity of a place cell $c_{i}$ should be 
represented by a vertex $v_{i}$; a \emph{detected} coactivity of two place cells, $c_{i}$ and $c_{j}$---by a 
bond $\sigma_{ij}$, a \emph{detected} coactivity of three place cells, $c_{i}$, $c_{j}$ and $c_{k}$---by a 
simplex $\sigma_{ijk}$ and so on (Fig.~\ref{Fig1}B). The resulting complex $\mathcal{T}_{\textrm{eff}}$ then 
constitutes a basic phenomenological model of a cognitive map assembled from the spiking inputs transmitted 
through imperfect synaptic connections.

%%%%%%%%%%%%%%%%%%%%%%%%%%%%%%%%%%%%%%%
%%%%%%%%%%%%%%%%%%%%%%%%%%%%%%%%%%%%%%%

\textbf{Statistical approach}. The mechanisms of spike generation, transmission and detection are probabilistic 
in nature. Transmitting action potentials requires producing a sufficient number of synaptic contacts in suitable 
locations of the postsynaptic neuron's membrane, releasing proper amount of neurotransmitter at each synapse at 
suitable times, inducing the excitatory postsynaptic potential (EPSP) of required magnitudes, etc., all of which 
involve probabilistic mechanisms \cite{Soltani,London}. In addition, there may appear flaws and glitches in the 
axons, synaptic clefts and in the structure of the postsynaptic membrane's polarization. Thus, there exists a 
probability $p_{k} < 1$ that a $k^{\mathrm{th}}$ connection in a cell assembly $\varsigma$ will induce sufficient 
EPSP in the readout neuron's membrane and a probability $q_{\varsigma} < 1$ that the latter will spike upon 
receiving the inputs (Fig.~\ref{Fig3}C).

In principle, these values could be estimated from the synaptic configuration of each individual assembly, which, 
however, would present a tremendous computational challenge \cite{Branco,Arleo,Garrido}. In order to avoid such 
complications, we will assume a basic statistical approach. First, we will regard the probabilities $p_{k}$, and 
$q_{\varsigma}$ as the prime parameters that describe the synaptic connections with the readout neuron. 
Second, we will view $p_{k}$ and $q_{\varsigma}$ as random variables, distributed according to a unimodal 
distribution, $P(p|\hat p,\Delta_{p})$ and $Q(q|\hat q,\Delta_{q})$ were $\hat p$ and $\hat q$ are the modes 
(the characteristic values) and $\Delta_{p}$ and $\Delta_{q}$ define the corresponding variances. 
Third, we will disregard synaptic plasticity processes and assume that the distributions are stationary, i.e., that 
the modes and the variances are fixed. Fourth, we will assume that both variables are distributed lognormally, as 
suggested by experimental observations \cite{BuzMi1,Barbour,Brunel}. We will also define the variances as functions 
of the modes, $\Delta_{p} \propto \hat p^2$ and $\Delta_{q} \propto \hat q^2$, which will allow us to exclude 
non-biological statistics and to study the topological properties of the emerging cognitive maps as functions of just 
two parameters, $\hat p$ and $\hat q$. 

\textbf{Implementation}. In order to isolate the effects of varying transition probabilities while keeping the temporal 
structure of the presynaptic spike trains ``clamped," we use the spiking data that was precomputed for the ``ideal" 
synaptic connections ($p_{k}=q_{\varsigma}=1$), and then screen out some of the spikes, to match each individual 
transmission probabilities $p_{k} < 1$ and to simulate the readout neurons' responses to the igniting cell assemblies 
with probabilities $q_{\varsigma} < 1$. 

To evaluate the latter, we reasoned as follows. Since in our approach the cell assemblies are modeled as the cliques 
of the coactivity graph $\mathcal{G}$, i.e., as composite objects assembled from $n(n-1)/2$ pairs of place cells, the 
probabilities of igniting the higher order place cell combinations can be computed from the pairwise coactivities. Indeed, 
if the spikes produced by the place cells $c_{i}$ and $c_{j}$ are transmitted to the readout neuron with the probabilities 
$p_{i}$ and $p_{j}$ respectively, then the corresponding pairwise coactivity occurs with the probability $p_{i} p_{j}$. 
The probability of a third order coactivity, e.g., the ignition of a clique $\sigma_{ijk}  = [c_{i}, c_{j}, c_{k}]$ is then 
defined by the probability of transmitting the coactive pairs $\sigma_{ij}=[c_{i},c_{j}]$, $\sigma_{jk}=[c_{j},c_{k}]$, 
and $\sigma_{ik} = [c_{i}, c_{k}]$ and detecting the result with the probability $q_{\varsigma}$; the probability of 
igniting the fourth order cliques is defined by the corresponding six coactive pairs and so forth.

With these assumptions, one can test how the spike transmission and detection probabilities affect the emergence of a 
spatial map, e.g., how synaptic depletion affects spatial learning, how the learning times and the topological structure of 
the cognitive map depend upon the strengths of synaptic connections between the place cells and the readout neurons, 
at what point spatial learning may fail, and so on.

\section{Results}
\label{section:results}

\begin{figure} [!ht]
	{\includegraphics[scale=0.80]{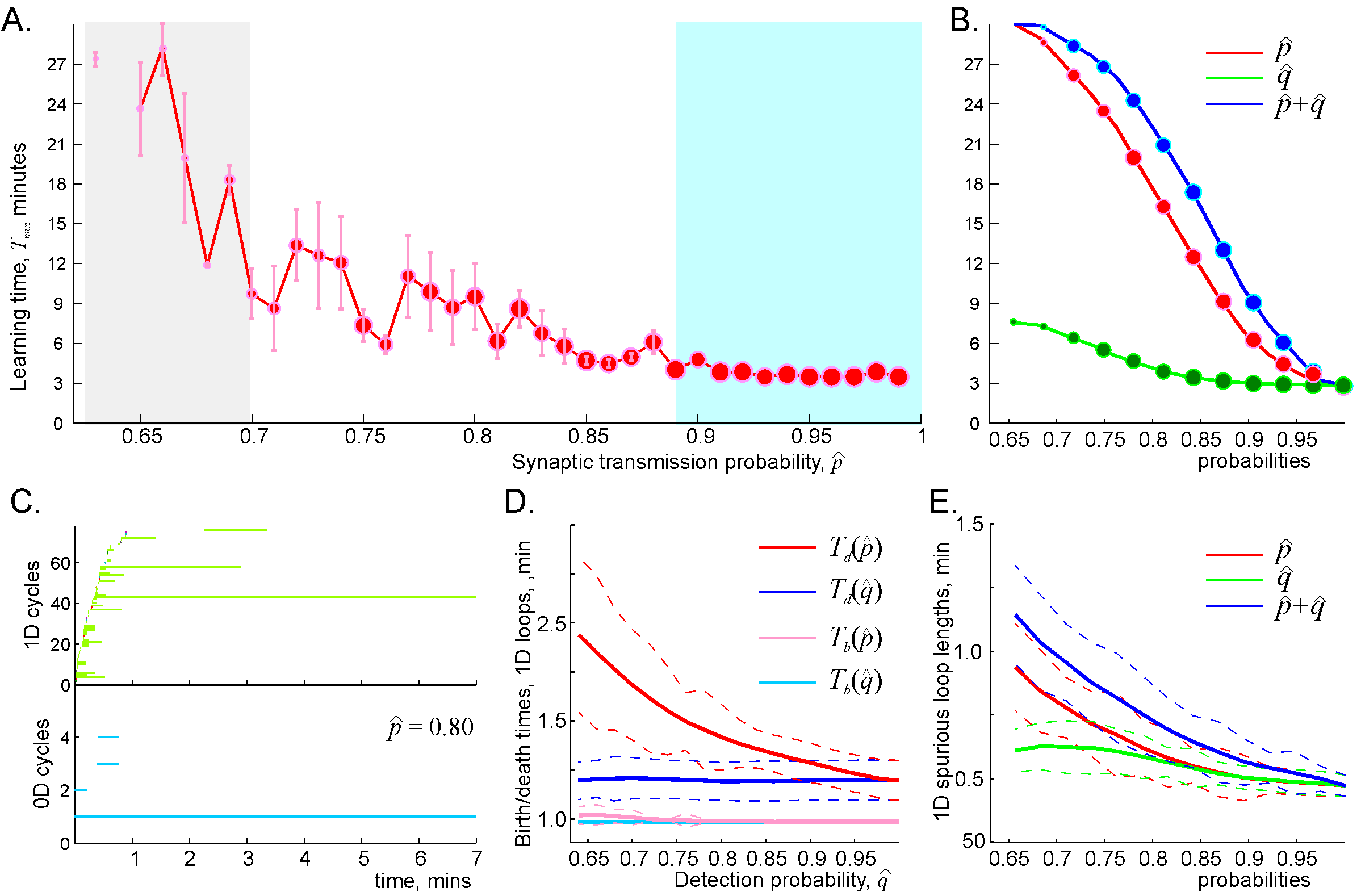}}
	\caption{\label{Fig4}
{\footnotesize
	\textbf{Synaptic transmission probability and the learning times}. ({\bf A}). The dependence of the learning 
	time $T_{\min}$ on the ensemble mean spike transmission probability, $\hat p$, in an ensemble of $N=400$ neurons 
	with a mean firing rate of $f = 28$ Hz, and mean place field size $30$ cm. The learning times, 
			$T_{\min}$, are computed for 40 values of $\hat p$, ranging between $\hat p = 1$ and $\hat p = 0.6$. 
			The size of the data points represents the percentage of the outcomes with the correct Betti numbers 
			($b_0(\mathcal{T}_{\textrm{eff}}) = b_1(\mathcal{T}_{\textrm{eff}}) = 1$). For high probability of 
			spike transmissions ($\hat p > 90\%$, blue-shaded area) the learning time remains nearly unchanged; as 
			$\hat p$ drops further, the learning time increases at a power rate. As the transmission probability 
			approaches the critical value $\hat p_{\textrm{crit}}$ (in this case, $\hat p_{\textrm{crit}}\approx 0.64$, 
			gray-shaded area), the learning times become large and highly variable; below $\hat p_{\textrm{crit}}$ the 
			place cell ensemble fails to form the correct topological map, even though the place cells exhibit perfectly 
			functional, spatially specific firing pattern.
			({\bf B}). The effect produced by the decreasing postsynaptic response probability ($\hat q$, green curve) 
			is similar, but smaller than the effect produced by the decreasing spike transmission probability ($\hat p$, 
			red curve). The combined effect (blue curve) is approximately additive, dominated by $\hat p$-dependence.
			({\bf C}). Timelines of $0D$ (blue) and $1D$ (green) topological loops computed for the same map and 
			$\hat p = 0.8$. This panel serves as an illustration for the next two panels. 
			({\bf D}). On average, the spurious loops appear in about a minute after the onset of the navigation, which 
			approximately corresponds to the time required to run around the central hole of the environment (Fig.~\ref{Fig1}A). 
			As the probabilities $\hat p$ or $\hat q$ decrease, the birth times ($T_{b}(\hat p)$ and $T_{b}(\hat q)$ the pink and 
			the light blue curve correspondingly) do not change significantly. In contrast, the times required by the spurious 
			loops to disappear grow significantly: $T_{d}(\hat p)$ (red curve) grows by over $100\%$, and $T_{d}(\hat q)$ 
			(blue curve) increases by a few percent.
			({\bf E}). The dependence of the spurious loops' length as a function of spike transmission $\hat p$, and the 
			readout neurons' response probability, $\hat q$.}
}
\end{figure}

\textbf{Learning times}. Lowering the characteristic probability of spike transmissions and the characteristic probability 
of the readout neurons' responses produces an uneven delay in spatial learning times (Fig.~\ref{Fig4}A). If the spike 
transmission probability is high (typically $0.9 \leq \hat p \leq 1$), then the small variations of $\hat p$ do not inflict a
strong impact on $T_{\min}$, i.e., the time required to learn the spatial map in a network with strong synaptic connections 
is nearly unaffected by occasional omissions of spikes. On the other hand, as $\hat p$ lowers to a certain critical value 
$\hat p_{\textrm{crit}}$, the learning times become high and, as $\hat p$ drops below $\hat p_{\textrm{crit}}$, the 
coactivity complex fails to produce the correct topological shape of the environment in finite time. For the intermediate 
values, the learning time increases at a power rate,
\begin{equation}
T_{\min} \propto (\hat p -\hat p_{\textrm{crit}})^{-\kappa},
\label{p}
\end{equation}
where $\kappa$ ranges between $0.1$ and $0.5$ for different values of $s,f,N$. The effects produced by the diminishing 
probability of the postsynaptic neurons' responses, $\hat q$, are qualitatively similar but weaker than the effects of 
lowering the spike transmission probability $\hat p$: the learning time shows a weak or no dependence for large $\hat q$ 
(typically $0.8\leq\hat q\leq 1$), followed by the power divergence near the critical value,
\begin{equation}
T_{\min} \propto (\hat q -\hat q_{\textrm{crit}})^{-\varkappa},
\label{q}
\end{equation}
with a small power exponent $\varkappa\approx 0.1$ (Fig.~\ref{Fig4}B). Lowering both $\hat p$ and $\hat q$ simultaneously 
leads to a combined, accelerated increase of the learning time (Fig.~\ref{Fig4}B). 

%%%%%%%%%%%%%%%%%%%%%%%%%%%%%%%%%%%%%%%
%%%%%%%%%%%%%%%%%%%%%%%%%%%%%%%%%%%%%%%

An implication of this phenomenon is that, $\hat p$ and $\hat q$, being independent characteristics of synaptic efficacy, can also 
compensate for each other's alterations: the effect of decreasing $\hat q$ can be counterbalanced by increasing $\hat p$ and vice 
versa. Indeed, the dependencies (\ref{p}) and (\ref{q}) also define the changes of the learning time induced by small variations in 
the transmission probability, 
\begin{equation}
\delta_{p}T_{\min}\propto -\frac{\kappa \, T_{\min}}{\hat p -\hat p_{\textrm{crit}}}\delta\hat p,
\end{equation}
and by the variations of the postsynaptic neuron's response probability, 
\begin{equation}
\delta_{q}T_{\min} \propto -\frac{\varkappa \, T_{\min}}{\hat q -\hat q_{\textrm{crit}}}\delta\hat q.
\end{equation} These relationships imply that the compensation of the changes of the learning time, 
$\delta_{p}T_{\min} = - \delta_{p}T_{\min}$, is achieved if 
\begin{equation}
\frac{\kappa \,\delta \hat p}{\hat p -\hat p_{\textrm{crit}}}\approx 
-\frac{\varkappa\,\delta\hat q}{\hat q -\hat q_{\textrm{crit}}}.
\end{equation} 
Notice, that this dependence is $T_{\min}$-independent and nonlinear: 
given a particular value of $\delta \hat p$, the required compensatory change of $\delta\hat q$ depends on the initial values of both 
$\hat p$ and $\hat q$.

\textbf{Dynamics of the effective coactivity complex}. The failures of the learning and memory capacity caused by deterioration 
of synapses are broadly discussed in the literature \cite{Selkoe,Neves,Mayford}. However, empirical observations provide only 
correlative links between these two scopes of phenomena. Indeed, the direct effects of the synaptic changes, e.g. the alterations 
of EPSP magnitudes, the spike transmission probabilities, the parameters of synaptic plasticity, etc., occur at cellular scale. It 
therefore remains unclear how such changes may accumulate at the network scale to control the net structure and the dynamics 
of the large-scale memory framework at the organismal level. The topological model allows addressing these questions at a 
phenomenological level, in terms of the structure of the coactivity complex $\mathcal{T}_{\textrm{eff}}$---its topological shape, 
its size, the dynamics of its topological loops and so forth, in response to the changes of synaptic parameters.

For example, one can evaluate the statistics of birth ($T_{b}$) and death ($T_{d}$) times of the topological loops in the 
coactivity complex. As shown on (Fig.~\ref{Fig4}C), the time when spurious loops begin to emerge depend only marginally on 
spike transmission probability. However, the spurious loops' disappearance times are impacted 
much stronger: although $T_{d}$ shows only weak $\hat p$-dependence at high $\hat p$, further suppression of 
the spike transmissions may double or triple the loops' disappearance time. The contribution of the decreasing 
response probability $\hat q$ is similar, but at a smaller scale: over the range $\hat q_{\textrm{crit}}<\hat q\leq 1$, 
the learning time changes only by a few percent (Fig.~\ref{Fig4}D). Similar effects are indicated by the $\hat p$- 
and $\hat q$-dependencies of the spurious loops' lengths, which may grow significantly as a result of the diminishing 
spike transmission probability, but increase only by $30-50\%$ due to the lowering probability of the readout neuron's 
responses (Fig.~\ref{Fig4}E). 

Taken together, these results explicate the power growth of the learning times indicated by (\ref{p}) and (\ref{q}) 
and provide a simple intuitive explanation for the decelerated spatial learning and its eventual failure caused by the 
synaptic depletion: according to the model, lowering synaptic efficacy stabilizes spurious topological loops in the 
coactivity complex, making it harder to extract physical information from the transient noise.

%%%%%%%%%%%%%%%%%%%%%%%%%%%%%%%%%%%
\begin{figure}[!ht]
	\includegraphics[scale=0.86]{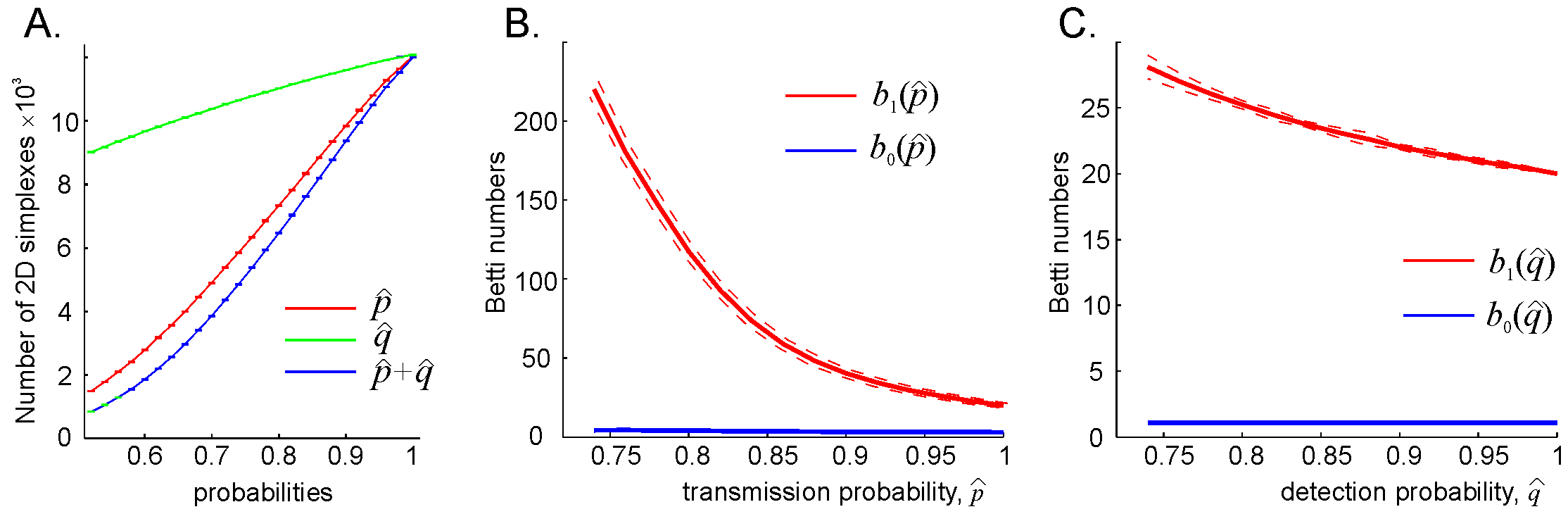}
	\caption{\label{Fig5} 
	{\footnotesize
	\textbf{Deterioration of the coactivity complex}. 
		({\bf A}). The size of the complex shrinks with the diminishing spike transmission ($\hat p$-dependence, red 
		line) and the readout neurons' response ($\hat q$-dependence, green line) probability at a power rate, $N_{2}
		\propto (\hat p-\hat p_{\textrm{crit}})^{\delta}$, and $N_{2} \propto (\hat q-\hat q_{\textrm{crit}})^{\epsilon}$. 
		In this case, $\delta \sim 1.5$ and $\epsilon \sim 0.4$. The combined effect of reducing both $\hat p$ and 
		$\hat q$ is illustrated by the blue line. 
		({\bf B}). The spurious topological loops in $\mathcal{T}_{\textrm{eff}}$ proliferate exponentially with 
		decreasing transmission probability $\hat p$. The blue and the red curve show the dependence of zeroth and 
		first Betti numbers on the transmission probability, $b_0(\hat p)$ and $b_1(\hat p)$ respectively.
		({\bf C}). The dependence of the numbers of $0D$ and $1D$ on the readout neuron's response probability 
		$\hat q$ is weaker: while $b_1(\hat q)$ exhibits a moderate growth, the $b_0(\hat q)$ remains unchanged.}
	}
\end{figure} 
%%%%%%%%%%%%%%%%%%%%%%%%%%%%%%%%%%%

Additional perspective on the mechanisms of the cognitive map's deterioration is produced by analyzing the size of the 
coactivity complex and the number of the topological loops in it. As shown on Fig.~\ref{Fig5}A, the decay of $\hat p$ 
causes rapid decay of the coactivity complex's size: the number of its two-dimensional simplexes (i.e., links in the 
coactivity graph, see below) drops as $N_{2} \propto (\hat p - \hat p_{\textrm{crit}})^{\delta}$, where $\delta > 1$. 
Diminishing $\hat q$ also shrinks the coactivity complex, but at a slower rate, $N_{2}\propto (\hat q -\hat 
q_{\textrm{crit}})^{\epsilon}$ with $0<\epsilon<1$. 
However, despite the shrinking size of the coactivity complex, the number of $1D$ spurious loops in it grows exponentially, 
$\log(b_1)\propto(\hat p_{\textrm{crit}} - \hat p)$, from a few dozen to a few hundred, accompanied by a weak $b_0
(\hat p)$ increase (Fig.~\ref{Fig5}B). Similar effects are produced by the lowering detection probability, but again, 
at a much smaller scale: the number of $1D$ loops, $b_1(\hat q)$, increases by about $30\%$ while the $b_0(\hat q)$
does not change (Fig.~\ref{Fig5}C).

These outcomes indicate that, as a result of weakening synaptic connections, the spurious topological loops do not only 
stabilize but also proliferate, thus preventing the effective coactivity complex from capturing the correct topology of the 
ambient space. In physiological terms, the model predicts that weakening synapses produce large numbers of longer-lasting 
topological defects in the cognitive map, which results in a rapid increase of the time required to learn the topology of the 
physical environment from poorly communicated spiking inputs. 

\textbf{Critical probabilities}. As indicated above, if the synaptic efficacies are too weak, i.e., if either the spike transmission 
or the postsynaptic response probability drops below their respective critical values, then the effective coactivity complex 
$\mathcal{T}_{\textrm{eff}}$ may disintegrate into a few disconnected pieces and lose its physical shape---a single large 
piece with a hole in the middle ($b_0(\mathcal{N})=b_0(\mathcal{E})$ and $b_1(\mathcal{N})=b_1(\mathcal{E})$, Fig.~\ref{Fig1}C), 
may be replaced by a ``spongy" configuration containing several smaller pieces with many holes \cite{Ambjorn,Hamber}. Thus, 
the cognitive map may appear in two distinct states: for $\hat p > \hat p_{\textrm{crit}}$ and $\hat q > \hat q_{\textrm{crit}}$ 
the spurious topological defects can be separated from the topological signatures of the physical environment, whereas below 
the critical values, topological noise overwhelms physical information. The transition between these two states is accompanied 
an increased variability of the learning times (Fig.~\ref{Fig4}A) and by their power divergence caused by the exponential 
proliferation of the topological fluctuations in the coactivity complex. These effects suggest that, near $\hat p_{\textrm{crit}}$ 
and $\hat q_{\textrm{crit}}$, the coactivity complex may experience a phase-like transition \cite{Franzosi1,Franzosi2,Vaccarino} 
from a regular state, in which spatial learning is effective to an irregular state, in which spatial learning fails.

Since in most of the studied cases, the critical synaptic transmission probability, $\hat p_{\textrm{crit}}$, is easier to achieve 
than the critical probability of the readout neuron's responses, $\hat q_{\textrm{crit}}$, we studied the dependence of the 
former on the ensemble parameters, i.e., on the number of place cells in the ensemble, their mean firing rate and the mean 
place field size, 
\begin{equation}
\hat p_{\textrm{crit}} = \hat p_{\textrm{crit}}(s,f,N).
\end{equation}

%%%%%%%%%%%%%%%%%%%%%%%%%%%%%%%%%%%
\begin{figure}[!ht]
	\includegraphics[scale=0.82]{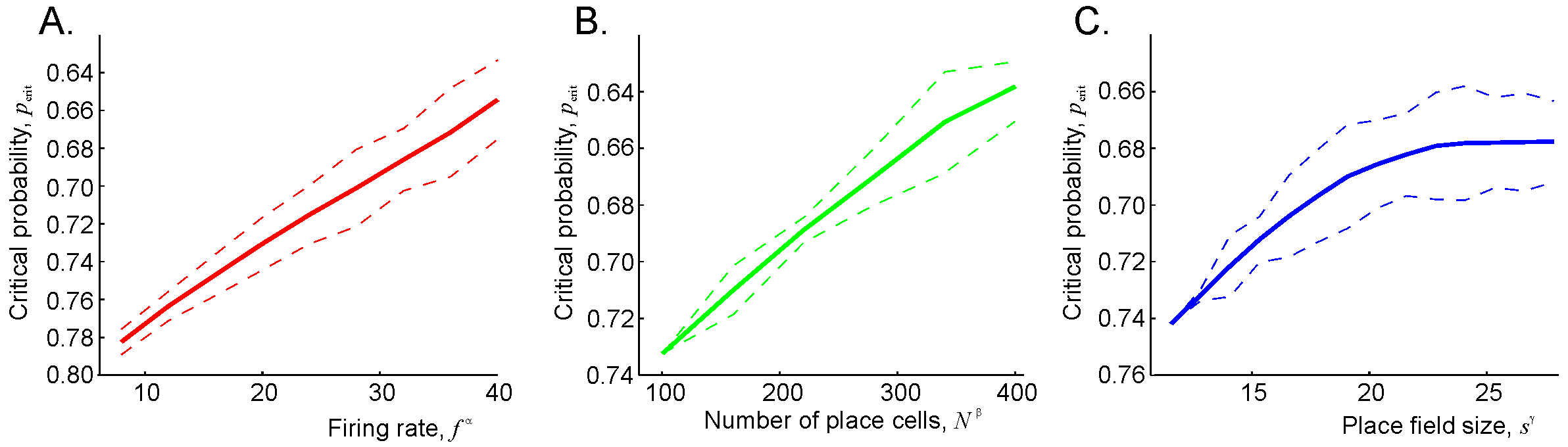}
	\caption{\label{Fig6} 
	{\footnotesize
	\textbf{Critical transmission probability as a function of spiking parameters}. 
		({\bf A}). Increasing the mean ensemble firing rate $f$ reduces the critical transmission probability at a superlinear 
		rate. Shown is the dependence $\hat p_{\textrm{crit}}$ vs. $f^{-\alpha}$, where $\alpha\approx 4$.
		({\bf B}). As the number $N$ of active place cell in the ensemble grows, the critical transmission probability drops 
		as $\hat p_{\textrm{crit}}\propto N^{\beta}$, with $\beta\approx 1.2$. 
		({\bf C}). The transmission probability also drops as a function of the mean place field size $s$, $\hat p_{\textrm{crit}}
		\propto s^{\gamma}$, $\gamma \approx 0.4$, as long as the place fields are not too large. As the place cells loose 
		spatial specificity of firing activity ($s>20$), a low transmission probability cannot be sustained.}
	}
\end{figure} 
%%%%%%%%%%%%%%%%%%%%%%%%%%%%%%%%%%%

The results shown on Fig.~\ref{Fig6} reveal power-law dependencies: $\hat p_{\textrm{crit}}\propto f^{\alpha}$, $\hat p
_{\textrm{crit}}\propto N^{\beta}$, and a more complex $s$-dependence. Since the domain of these dependences covers the 
experimentally observed range of parameters, the results can be interpreted physiologically. First, if the ensemble firing rates are 
too low, or if the place fields are too meager, or the number of the active neurons is too small (the left ends of the dependencies 
shown on Fig.~\ref{Fig6}), then the corresponding place cell ensemble fails to learn the spatial map of the environment, even if 
the synaptic connections are nearly perfect ($\hat p  > 0.75$), which corresponds to the results discussed in \cite{PLoS,Arai,Basso}.
As the mean firing rate and the number of active neurons increase, the critical probability $\hat p_{\textrm{crit}}$ steadily 
\emph{decreases}, which implies that the synaptic depletion may be compensated by enhancing neuronal activity, as observed in 
experimental studies \cite{Palop,Hartley,Busche}. In contrast, the dependence $\hat p_{\textrm{crit}}(s)$ saturates and even 
reverses its direction for overly large place fields. This, however, is a natural result since poor spatial specificity of the place cells' 
spiking should prevent successful spatial leaning even for large $\hat p$ \cite{PLoS,Arai,Basso}.

Electrophysiological studies show that only up to $10-20\%$ of spikes are transmitted between the neurons in CA1 slices, which is 
lower than the critical values discussed above \cite{BuzMi1,Csicsvari,BuzMi2}. However, the results shown on Fig.~\ref{Fig6}C imply 
that the experimentally observed values of $\hat p$  can be achieved for larger values of $N$, i.e., in larger place cell ensembles. 
Interpolated $\hat p_{\textrm{crit}}(N)$ dependence indicates that the physiological values of $\hat p_{\textrm{crit}}\sim 0.1-0.2$, 
can be achieved for the ensembles of $N\sim 3000$ cells, which corresponds to the experimentally observed values \cite{Wilson1,Leutgeb1}.

\textbf{Learning region}. One of the key characteristics of the place cell spiking activity produced by the topological model is the 
range of the spiking parameters, for which the coactivity complex can assume a correct topological shape in a biologically feasible 
period. Geometrically, this set of parameters forms a domain in the parameter space that we refer to as the \emph{learning region}, 
$\mathcal{L}$ \cite{PLoS}. The shape and the size of the learning region varies with the geometric complexity of the environment and 
the difficulty of the task: the simpler is the environment and easier the task, the larger is $\mathcal{L}$, i.e., the wider the range 
of physiological values that permits learning a map of that space \cite{Nithianantharajah,Hernan}. On the other hand, a larger $\mathcal{L}$ 
implies a greater range within which the brain can compensate for physiological variation: if one parameter begins to drive the system 
outside the learning region, then successful spatial learning can still occur, provided that compensatory changes of other parameters 
can keep the neuronal ensemble inside $\mathcal{L}$. For example, a reduction of the number of active neurons can sometimes be 
compensated by adjusting the firing rate or the place field size in such a way as to bring their behavior back within the perimeter of 
the learning region.

Interpreting the parameters of a given place cell ensemble in the context of its placement within or relative to the learning region 
sheds light on the mechanism of memory failure caused by certain neurophysiological conditions, e.g., by the Alzheimer Disease 
\cite{Cacucci,LaFerla}, or by aging \cite{Robitsek,Wilson2} or certain chemicals, e.g., ethanol \cite{White,Matthews}, cannabinoids 
\cite{Robbe1,Robbe2} or methamphetamines \cite{Kalechstein,Silvers}, which appear to disrupt spatial learning by gradually shifting 
the parameters of spiking activity beyond the learning region. On the other hand, the performance of a place cell ensemble can improve 
by enhancing place cells spiking activity pharmacologically or by Deep Brain Stimulation \cite{Laxton}, or by modulating the hippocampal 
neural oscillations \cite{Shirvalkar}, as the model predicts \cite{PLoS,Arai,Basso}.

\begin{figure}[!ht]
	\includegraphics[scale=0.86]{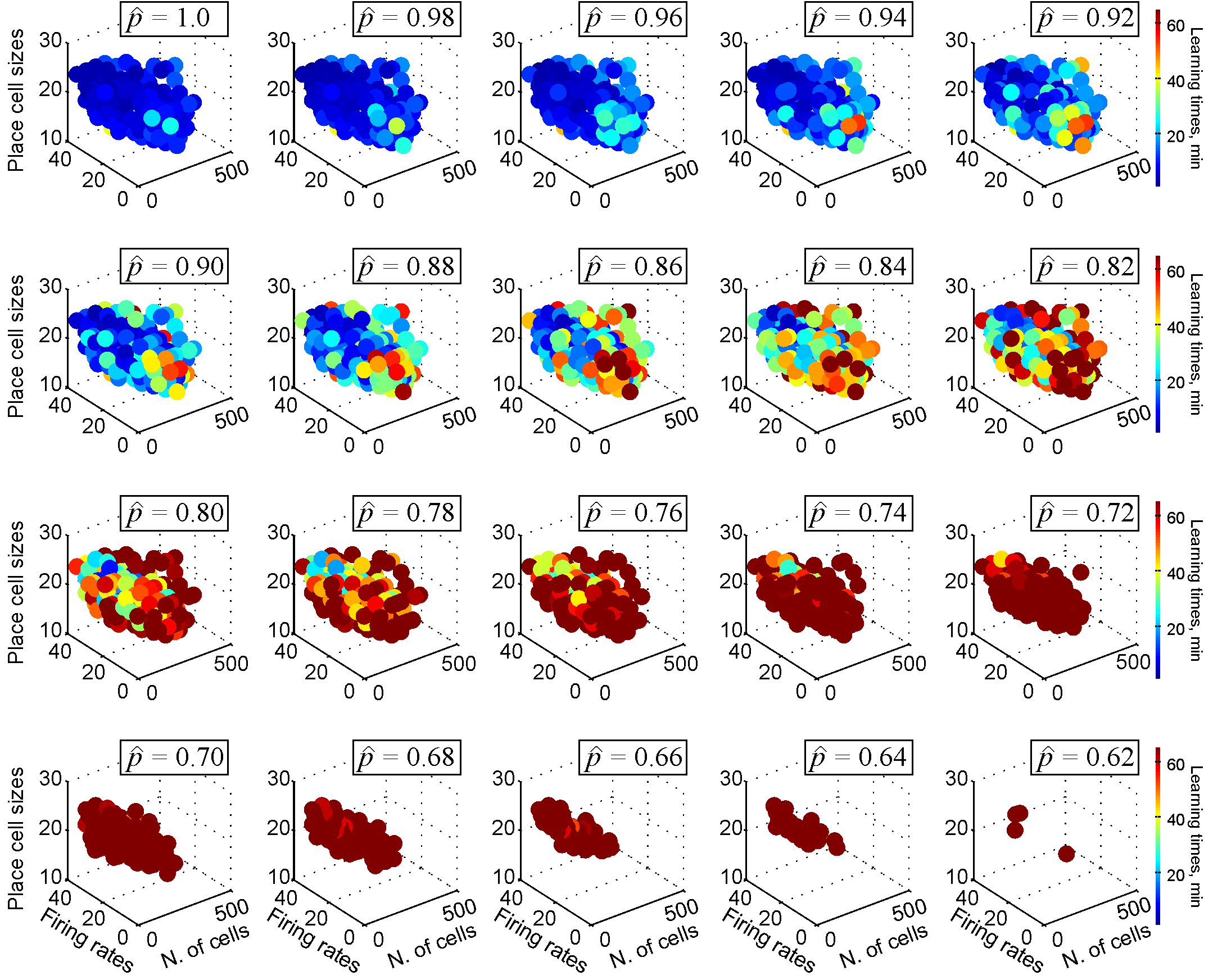}
	\caption{\label{Fig7} 
	{\footnotesize
	\textbf{Synaptic connection strengths affect spatial learning}. By simulating 
		spatial learning in a given environment for various ensemble-mean values of place field size 
		($4 < s < 30$), firing rates ($4 < f < 40$) and the size of the place cell population 
		($50 < N < 500$), we can estimate the domain within a large parametric space representing the set 
		of place cell ensembles that can produce a correct spatial map---the \emph{Learning Region} 
		($\mathcal{L}$). Each dot represents a hippocampal state as defined by a particular triple $(s, f, N)$; 
		the color of the dot is the mean time required for a given ensemble to encode an accurate map of the 
		environment's features, averaged over ten place field configurations. Outside $\mathcal{L}$ learning 
		is inaccurate or unreliable. As the transmission probability $\hat p$ decreases, the learning region 
		shrinks and disappears as the transmission probability $\hat p$ approaches $\hat p_{\textrm{crit}}$.}
	}
\end{figure}

In contrast, diminishing spike transmission probability produces a qualitatively different effect: as shown on Fig.~\ref{Fig7}, it 
reduces the learning region from its original (largest) size at $\hat p = 1$ to its compete disappearance at the critical value $\hat p 
= \hat p_{\textrm{crit}}$. During this process, the time required to form the cognitive map of the environment progressively increases 
from a few minutes to over an hour (Fig.~\ref{Fig7}).

Physiologically, these results suggest that if the synaptic connections are too weak, then the system may fail to form a map not only 
because the parameters of neuronal firing are pushed beyond a certain ``working range," but also because that range itself may diminish 
or cease to exist. In particular, the fact that the learning region disappears if the transmission probability drops below the critical 
value, implies that the deterioration of memory capacity caused by synaptic failure may not be compensated by increasing the place field's 
firing rates or by recruiting a larger population of active neurons, i.e., some neuropathological conditions may indeed be primarily 
``synaptic'' in nature \cite{Selkoe}.

%%%%%%%%%%%%%%%%%%%%%%%%%%%%%%%%%%%
%%%%%%%%%%%%%%%%%%%%%%%%%%%%%%%%%%%

\textbf{Deteriorating cognitive graph} A simple alternative explanation of these results can be provided in terms of the place cell 
coactivity statistics. As pointed out in Section~\ref{section:model}, the collection of the unique pairs of the coactive place cells in a 
network with ideal synaptic connections ($\hat p = \hat q = 1$) is represented by the coactivity graph $\mathcal{G}$. 
The imperfect synapses diminish the pool of the transmitted and the detected coactive pairs, which then corresponds to a smaller, 
\emph{effective} coactivity graph $\mathcal{G}_{\textrm{eff}}(\hat p)\subset\mathcal{G}$. The corresponding set of higher order 
coactivities---the effective coactivity complex $\mathcal{T}_{\textrm{eff}}(\hat p)$ induced from $\mathcal{G}_{\textrm{eff}}(\hat p)$ 
is a subcomplex of the original coactivity complex, with potentially altered topological properties. The net results discussed above 
imply that, for high transmission probabilities, the effective coactivity complex $\mathcal{T}_{\textrm{eff}}$ retains the original 
topological shape of $\mathcal{T}$, but as $\hat p$ diminishes, the effective complex shrinks, acquires multiple topological defects 
and eventually loses its correct shape, indicating a failure of spatial learning.

An illuminating perspective on the changing structure of the coactivity graph $\mathcal{G}$ described above is provided by its 
Forman curvature---a combinatorial analogue of the standard differential-geometric notion of curvature \cite{Forman,Lewiner}.
The Forman curvature is adopted for discrete, combinatorial structures, such as datasets, networks and graphs \cite{Weber1,
Weber2,Weber3,Sreejith}, and can be flexibly defined in terms of an individual network's characteristics---the ``weights" of its 
vertexes and edges. Specifically, for an undirected edge $e$ with a weight $w(e)$ connecting the vertexes $v_{1}$ and $v_{2}$ 
with the weights $w(v_{1})$ and $w(v_{2})$ it is defined as
\begin{equation}
R_{F}(e)= w(v_{1}) + w(v_{2}) - \Sigma_{e_{v_{1},v_{2}}} \left(w(v_{1})\sqrt{\frac{w(e)}{w(e_{v_1})}} + w(v_{2})
\sqrt{\frac{w(e)}{w(e_{v_2})}}\right),
\label{Re}
\end{equation}
where the summation goes over the other edges $e_{v_1}$ and $e_{v_2}$ connecting to $v_1$ and $v_2$. The curvature associated with a 
vertex, $R_F(v)$, equals to the mean curvature of the edges that meet at $v$.

%%%%%%%%%%%%%%%%%%%%%%%%%%%%%%%%%%%
\begin{figure} 
	\includegraphics[scale=0.86]{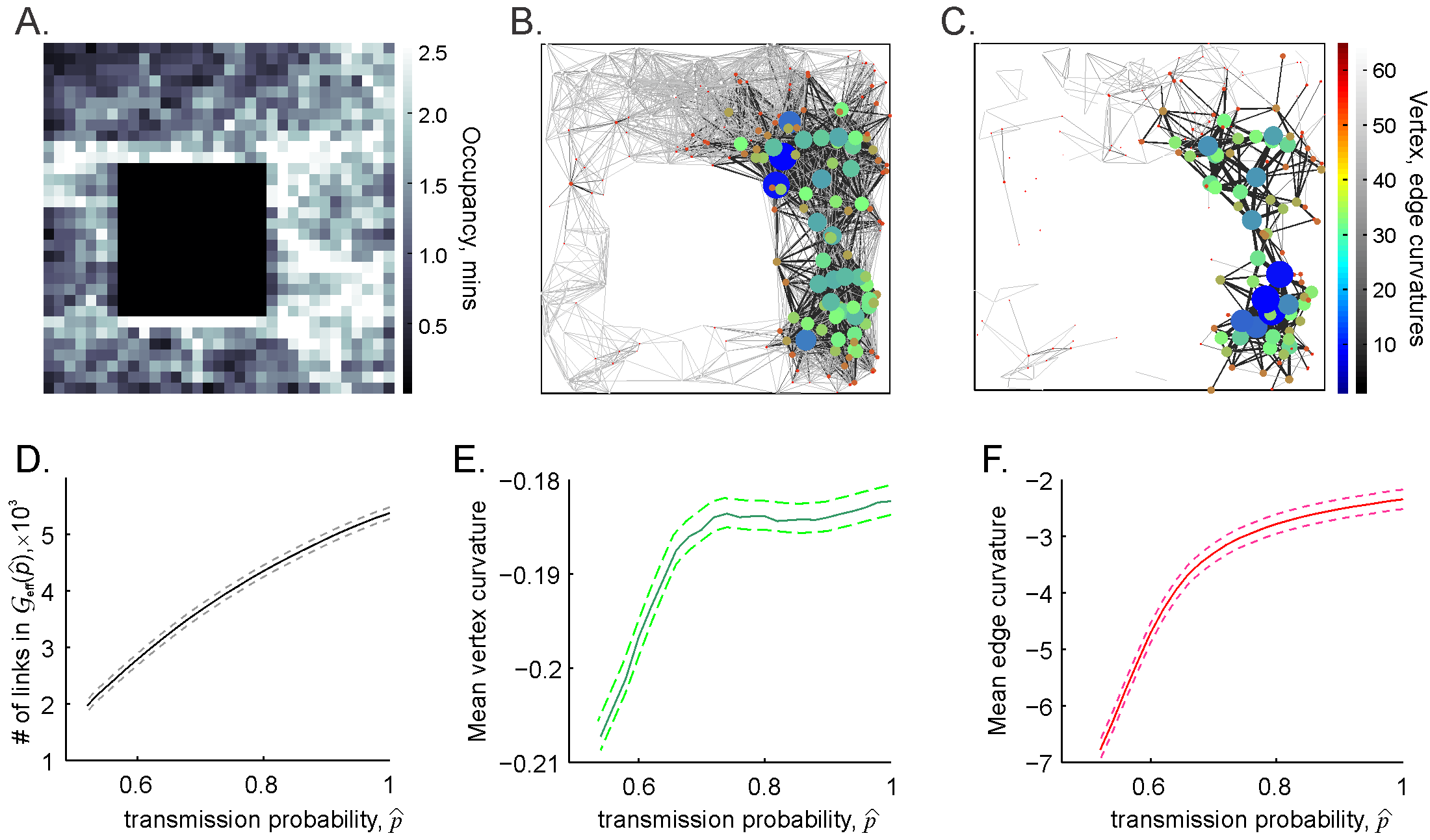}
	\caption{\label{Fig8} 
	{\footnotesize
	\textbf{The decay of the coactivity graph}. ({\bf A}). The occupancy map of the simulated rat's
		trajectory. The highlighted areas indicate where the rat spends more time (gray colorbar). ({\bf B}). The coactivity 
		graph $\mathcal{G}$ computed for perfect connections ($\hat p = \hat q = 1$): the thickness and the shade of the 
		edges (gray colorbar), as well as the sizes and the colors of the vertices (jet colorbar) are scaled according to 
		their respective Forman curvatures. ({\bf C}) The effective coactivity graph $\mathcal{G}_{\textrm{eff}}$ computed 
		for the spike transmission probability $\hat p = 0.65$ is significantly sparser that at $\hat p = 1$: only about $10\%$ 
		of edges with high Forman curvatures remain. Notice that in both cases, the edges and vertexes with high curvature 
		concentrate in the area that were visited most by the rat (panel A).
		({\bf D}) The effective coactivity graph shrinks as a function of the transmission probability decay. The mean Forman 
		curvature of the vertexes (panel {\bf E}) and of the edges ({\bf F}) also decreases as a function of decaying $\hat p$,
		as the low-curvature vertexes and edges disappear.}
	}
\end{figure} 
%%%%%%%%%%%%%%%%%%%%%%%%%%%%%%%%%%%

As discussed in \cite{Weber1,Weber2,Weber3,Sreejith}, the values $R_{F}(e)$ and $ R_{F}(v)$ provide a measure of the divergence 
of information flow across the network, highlighting the most ``important" edges and vertexes. Applying these ideas to the case 
of the coactivity graph, weighing its vertexes with the number of spikes produced by the corresponding place cells and its edges 
with correlation coefficients between the corresponding pairs of cells, reveals that the distribution of the resulting Forman 
curvatures follows the structure of the occupancy map (Fig.~\ref{Fig8}A,B). In other words, the most visited vertexes and edges 
appear as the most ``curved" ones, controlling the flow of information in $\mathcal{G}_{\textrm{eff}}$.

This quantification also allows a natural interpretation of the effective coactivity graph's dynamics: as the spike transmission 
probability $\hat p$ decreases, $\mathcal{G}_{\textrm{eff}}(\hat p)$ sheds the ``least important'' vertexes and links with low 
curvatures (Fig.~\ref{Fig8}C,D,E). Thus, as the synaptic efficacies weaken, the emerging effective coactivity graph reflects only 
the most persistently firing place cells and the highly correlated pairs of such cells, which can sustain the full topological 
connectivity information, but only for so long. As the synapses deteriorate below critical value, $\hat p<\hat p_{\textrm{crit}}$, 
the corresponding effective coactivity complex acquires an irreparable amount of topological defects and fails to encode the correct 
topological map of the environment.

\section{Discussion}
\label{section:discussion}

Countless observations point out that deteriorations of synapses often accompany memory deficiencies. For example, the recurrent 
connectivity of CA3 area of the hippocampus and the many-to-one projections from the CA3 to the CA1 area \cite{Shepherd,Syntax} 
suggest that the CA1 cells may provide readouts for the activity of the CA3 place cell assemblies \cite{CAs}. Behavioral and 
cognitive experiments demonstrate that weakening of the synapses between these two areas, a reduction in the number of active 
neurons in either domain, diminishing neuronal activity and so forth, correlate with learning and memory deficiencies observed, 
e.g., in Alzheimer's disease \cite{Cacucci,LaFerla} or in aging subjects \cite{Robitsek,Wilson2}. However, without a theoretical 
framework that can link the ``synaptic" and the ``organismal" scales, the detailed connections between these two scopes of 
phenomena are hard to trace. For example, if the spike transmission rate in an ensemble of place cells decreases, e.g., by $5\%$, 
will the time required to learn the environment increase by $1\%$, $5\%$ or by $50\%$? 
Does the outcome depend on the ``base" level of the transmission probability? Can an increase in learning time caused by 
synaptic depression always be compensated by increasing the population of active cells, or by elevating their spiking rates? 
The topological model permits addressing these questions computationally, at a phenomenological level, thus allowing us 
to move beyond mere correlative descriptions to a deeper understanding of the spatial memory deterioration mechanisms.

The hypothesis about topological nature of the hippocampal map \cite{eLife} is broader than the proposed Algebraic Topology (AT) 
model or the scope of questions that this model allows addressing. For example, the description based on the AT algorithms does 
not capture biologically relevant metrical differences between topologically equivalent paths or qualitative differences between 
topologically equivalent environments, e.g., between the widely used \textsf{W}-, \textsf{U}- or \textsf{T}-mazes, even though such 
differences are reflected in the place cell spiking patterns and are known to affect animals behavior \cite{Frank,Pastalkova,Jadhav}. 
Addressing these differences requires using alternative mathematical apparatuses, e.g., Qualitative Space Representation (QSR) 
techniques, such as Region Connection Calculi (\textsf{RCC}) \cite{Cui,Hazarika,Cohn}, which would complement the scope of topological 
methods used in neuroscience \cite{SchemaS}. In the current approach, we use AT instruments to assess a particular scope of questions, 
namely to estimate the conditions that guarantee structural integrity of the cognitive map and to describe its overall topological shape. 

%%%%%%%%%%%%%%%%%%%%%%%%%%%%%%%%%%
\begin{wrapfigure}{c}{0.5\textwidth}
	%\centering
	\includegraphics[scale=0.43]{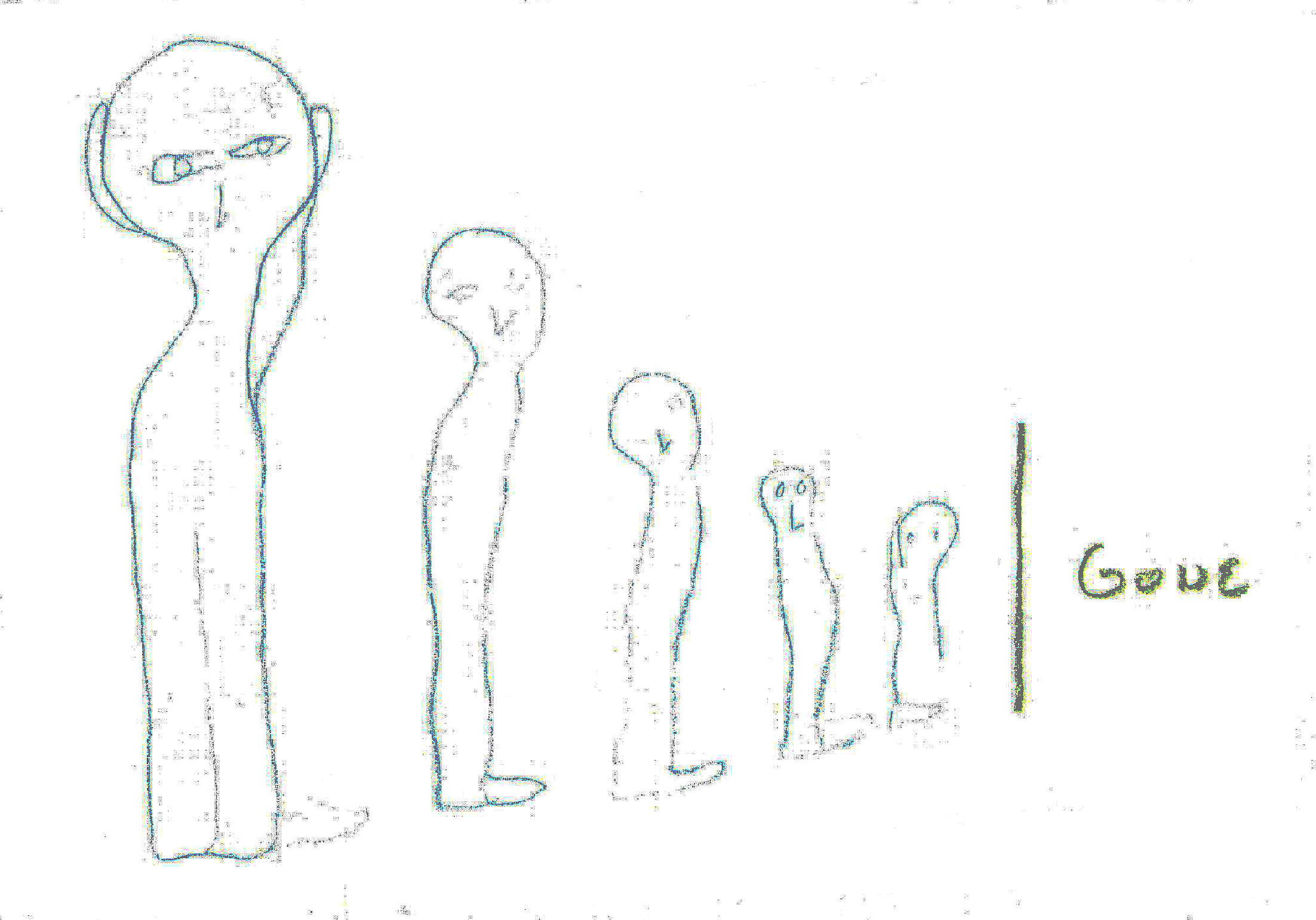}
	\caption{\label{Fig9} 
		{\footnotesize 
			A drawing by an Alzheimer disease patient depicting his own perception of the disease's 
			development. The word to the right of the vertical bar is ``Gone".}
	}
\end{wrapfigure}
%%%%%%%%%%%%%%%%%%%%%%%%%%%%%%%%%%%

Fundamentally, producing a cognitive map requires two key components: a proper temporal structure of the spike trains and a physiological 
mechanism for detecting and interpreting neuronal coactivity---a suitable network architecture, a proper distribution of the connectivity 
strengths, of the parameters of synaptic plasticity, etc. All these components influence spike transmission and detection probabilities, 
which, in our model, affect the shape and topological structure of the coactivity complex, the statistics of learning times, the structure 
of the learning region, etc. This produces a quantitative connection between the information processed at the microscopic level (neurons 
and synapses) and the properties of the large-scale representations of space emerging at the organismal level, described here by means of 
the Persistence Homology theory \cite{Zomorodian,GhristBar,Edelsbrunner}.

Lastly, it should be pointed out that the low-dimensional components of the coactivity complexes were used above to represent cognitive maps, 
i.e., frameworks spatial memories. However, the combinations of the coactive place cells, modeled as simplexes of $\mathcal{T}_{\textrm{eff}}$, 
may represent generic memory elements \cite{Wood,Ginther}. 
In other words, it can be argued that the net structure of $\mathcal{T}_{\textrm{eff}}$ represents not only spatial, but also nonspatial 
memories---a larger memory framework that can be viewed as a ``memory space" \cite{Eichenbaum1,SchemaM}. Thus, a disintegration of the 
cell assembly complex caused by deteriorating synapses discussed above may also be viewed as a model of the full memory space decay. 
From such perspective, it may be noticed that the results of the model parallel the experience of patients acquiring a slowly progressing 
dementia. For example, the model provides an explanation for the reason cognitive declines often do not manifest until quite a lot of 
damage has occurred. It also predicts that when the weakening synapses deteriorate beyond the range of parameters within which learning 
is effective, the damages push the neuronal ensemble beyond the bounds of the learning region. As a result, the failure becomes more 
frequent, and finally, the brain cannot perform that particular learning task, certain memories or abilities begin to flicker and then 
are lost mostly for good (Fig.~\ref{Fig9}). 

\section{Methods}
\label{section:methods}

The computational algorithms used in this study were described in \cite{PLoS,Arai}:

\textbf{The simulated environment} shown on Fig.~\ref{Fig1}A is designed similarly to the arenas used in typical electrophysiological 
experiments. Combining such small arenas allows simulating learning in larger, more complex environments \cite{Arai}. The simulated 
trajectory represents non-preferential, exploratory spatial behavior, with no artificial patterns of moves or favoring of one segment 
of the environment over another.

\textbf{Place cell spiking} probability was modeled as a Poisson process with the rate 
\begin{equation}
\lambda_c(r)=f_c e^{-\frac{(r-r_c)^2}{2s^2_c}}
\nonumber
%\label{lambda}
\end{equation}
where $f_c$ is the maximal rate of place cell $c$ and $s_c$ defines the size of its place field centered at $r_c = (x_{c}, y_{c})$ 
\cite{Barbieri}. In an ensemble of $N$ place cells, the parameters $s_{c}$ and $f_{c}$, are log-normally distributed with the means 
$f$ and $s$ and the variances $\sigma_{f}$  and $\sigma_{s}$. To avoid overly broad or overly narrow distributions, we used additional 
conditions $\sigma_{f}  = af$ and $\sigma_{s} = bs$, with $a = 1.2$ and $b = 1.7$ \cite{PLoS}. In addition, spiking probability was 
modulated by the $\theta$-wave \cite{Arai,Mizuseki,Huxter}. The $\theta$-wave also defines the temporal window $w \approx 250$ ms (about 
two $\theta$-periods) for detecting the place cell spiking coactivity, as suggested by experimental studies \cite{Mizuseki,Huxter,BuzsakiTh} 
and by our model \cite{Arai}. This value also defines the timestep used in the computations. The place field centers $r_c$ for each computed 
place field map were randomly and uniformly scattered over the environment. 

\textbf{Place cell ensembles} are specified by a triple of parameters $(s, f, N)$ and hence the learning region represents a domain of this 
$3D$ parameter space. The ensembles studied above contain between $N = 50$ and $N = 400$ place cells. The ensemble mean peak firing rate $f$ 
ranges from $4$ to $40$ Hz, and the average place field size ranges between $\sim 12$ cm and $\sim 90$ cm ($4 \leq s \leq 30$ cm).

\textbf{Persistent Homology Theory} is used to describe the evolving topological shape of the coactivity complexes in terms of their 
homological invariants \cite{Zomorodian,GhristBar,Edelsbrunner}. In particular, it allows computing the time-dependence of the Betti numbers and deducing 
the dynamics of its topological loops---their mean lifetimes, their mean lengths, their numbers, etc. Computations were performed using 
\textsf{javaplex} computational software developed at Stanford University \cite{JPlex}.

\section*{Acknowledgements}
\label{section:acknow}

The work was supported by the NSF 1422438 grant.

\section{References}
\label{section:refs}

\end{document}